\documentclass[aps,prd,twocolumn,amsmath,amssymb,amsfonts,nofootinbib,superscriptaddress,altaffilletter,showpacs,floatfix]{revtex4-1}
\usepackage{amsgen,amstext,amsbsy,amsopn,amsfonts,amssymb}
\usepackage{supertabular}
\usepackage[pdftex]{graphicx}
\usepackage{hyperref}
\hypersetup{bookmarksopen=true}
\usepackage{graphpap}
\usepackage{rotating}
\usepackage{url}
\usepackage{color}
\usepackage{dcolumn}

\def\Var{\textrm{Var\,}}

\def\E{\mathtt E}
\def\UL{\textrm{UL}}
\def\D{{\mathcal D}}

\def\expect{\mathbb E}

\definecolor{green4}{rgb}{0,0.25,0}

\def\dochecks{0}
\def\check#1{{\if \dochecks 1 {\color{green4}\ \\ \rule{0.4\linewidth}{1pt}\hfill Check \hfill\rule{0.4\linewidth}{1pt} \hspace*{0pt} \par #1 \rule{\linewidth}{1pt} } \fi}}

\begin{document}
\title{A Novel Universal Statistic for Computing Upper Limits in Ill-behaved Background}
\author{V.~Dergachev}
\date{\today}
\affiliation{
LIGO Laboratory,
California Institute of Technology,
MS 100-36,
Pasadena, CA 91125, USA
}
\begin{abstract}
Analysis of experimental data must sometimes deal with abrupt changes in the distribution of measured values. Setting upper limits on signals usually involves a veto procedure that excludes data not described by an assumed statistical model. We show how to implement statistical estimates of physical quantities (such as upper limits) that are valid without assuming a particular family of statistical distributions, while still providing close to optimal values when the data are from an expected distribution (such as Gaussian or exponential). This new technique can compute statistically sound results in the presence of severe non-Gaussian noise, relaxes assumptions on distribution stationarity and is especially useful in automated analysis of large datasets, where computational speed is important.
\end{abstract}
\maketitle

\section{Introduction}
Data collected in experiments are sometimes contaminated by noise or background with an ill-behaved and often unknown distribution, presenting problems for the traditional method of using distribution quantiles to establish upper limits or confidence intervals. This problem happens especially often in experiments that collect large volumes of data.

\begin{figure}[htbp]
\includegraphics[width=3.4in]{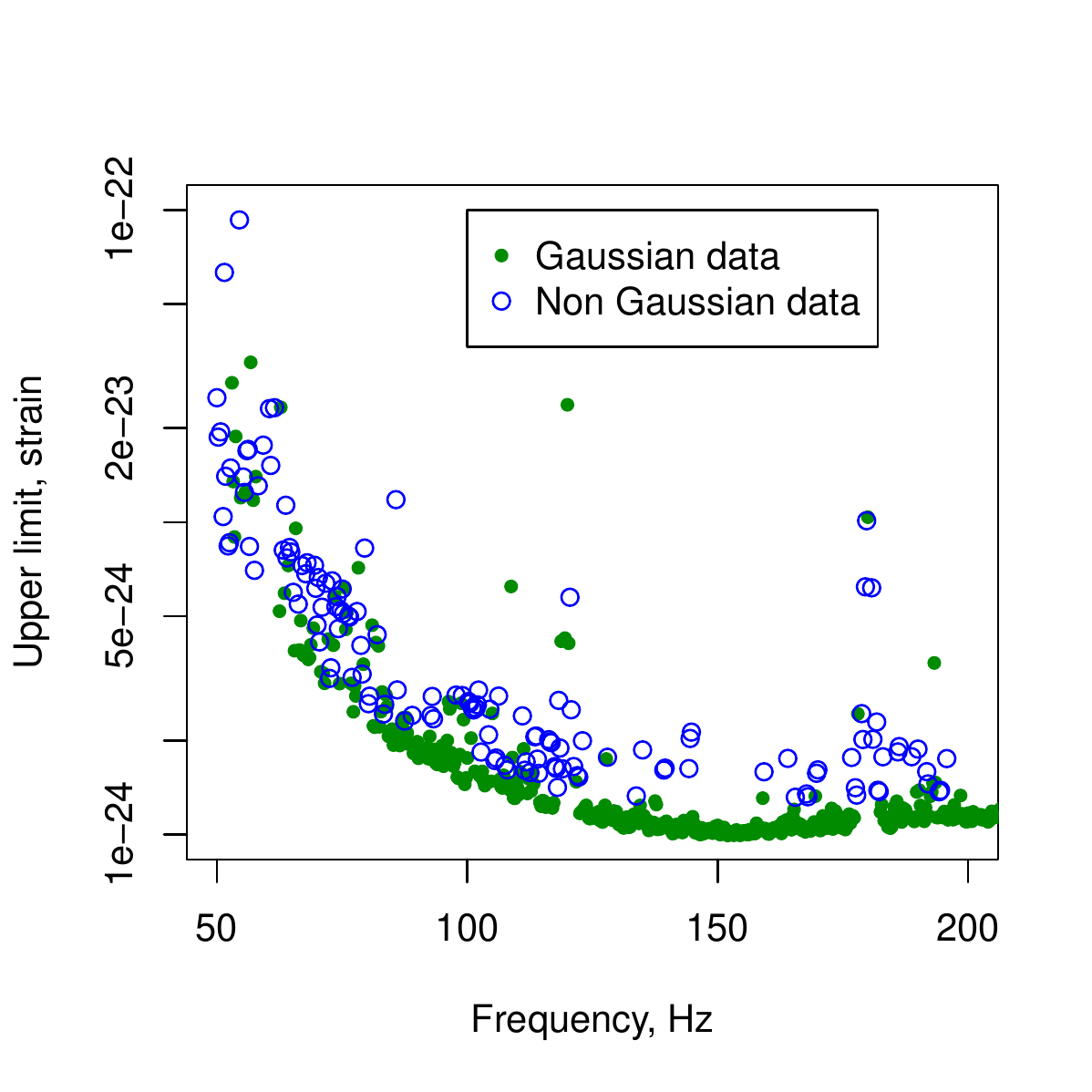}
\caption{Upper limit data from LIGO S5 search for continuous gravitational waves in the 50-200 Hz frequency range\cite{FullS5Semicoherent}. A large number of non-Gaussian bands significantly reduces the usefulness of the results. (color online)}
\label{fig:S5_50_200}
\end{figure}

A common solution is to exclude contaminated data from the analysis. For example, figure \ref{fig:S5_50_200} shows a small portion of data obtained in the LIGO search for continuous gravitational waves in fifth science run using PowerFlux code \cite{FullS5Semicoherent}. The blue points mark regions where non-Gaussian behavior was detected and upper limit values are not expected to be valid.

In fact, if one looks carefully at the data for each point one can find a cause of non-Gaussian behavior and a workaround to establish an upper limit - but the causes are different for different points, making analysis very laborious.

What is desired is an automated way to establish an upper limit that would be valid (if a bit conservative) for an arbitrary distribution, while still being close to optimum in the case of Gaussian noise (or other distribution class) that commonly occurs in the data. 

We present a new algorithm that establishes upper limits without assuming a specific underlying background distribution, and that can be optimized for an arbitrary class of distributions (such as Gaussian, exponential, etc) that are expected to commonly occur in the data. A comparison is made to conventional methods of establishing upper limits.

This advance allows one to obtain valid upper limits on signal strengths in the presence of ill-behaved and poorly understood background.

\section{Universal inequalities and statistic}

Let us consider a sample problem. Suppose we have obtained many samples of data which consist of background noise plus a possible signal. The data is collected in batches of $N$ samples $d_i$ for which the background noise $\xi_i$ is independent and identically distributed. Also, we expect that at most one sample $j$ in each batch contains a signal:
\begin{equation}
d_i=\xi_i+s \delta_{ij} 
\end{equation}
We would like to place a limit on the strength of the signals that may (or may not) be present in our data set.

If we knew that the noise $\xi_i$ were drawn from a particular distribution $\rho$ (such as Gaussian) our task would be straightforward - we would find the maximum $d_i$ in all our data, subtract the mean of the background $\mu$ and add a distribution specific correction $C_\rho$ that accounts for the possibility that a particular sample with the signal was below the background mean:
\begin{equation}
\UL_\rho=\max_i d_i - \mu + C_\rho
\end{equation}

If the distribution of $\xi_i$ is not known with certainty, then we can try to estimate the distribution from the data itself. This, however, is problematic when the amount of data is small. 

We now notice that, regardless of the procedure to compute the correction, it ends up being the function of the input data:
\begin{equation}
\UL=\max_i d_i - \mu + C(\left\{d_i\right\})
\end{equation}

We can now pose the following problem:

Suppose we are given a confidence level $1-\epsilon$, a class of commonly encountered distributions $\D$ and a tolerance $\alpha$. We need to find a function $C(\left\{d_i\right\})$ of the input data such that:
\begin{enumerate}
 \item For any $s$ and any distribution of $\xi_i$ we have
\begin{equation}
P(\UL<s)<\epsilon
\end{equation}
\item We require that for any distribution $\rho \in \D$ the upper limits are overestimated by at most $\alpha$ compared to what we could obtain with full knowledge of the distribution:
\begin{equation}
\expect \left(\frac{\UL}{\UL_\rho} \right)\le 1+\alpha 
\end{equation}
\end{enumerate}

We call such statistics {\em universal} as they are applicable regardless of the distribution of noise we have.

\section{Rank based upper limit statistic}

A universal statistic can be constructed by using quantiles of the data $\xi_i$. 

For example, to compute $95\%$ upper limit, one would find the maximum $M$ of the data $\xi_i$, and the $N/20$ smallest sample $V$. Then the upper limit is computed as
\begin{equation}
\UL_{\textrm{quantile}}=M-V
\end{equation}

This well-known procedure works quite well, except for small sample sizes $N<20$. 

For large scale computation we run into difficulties: while $A N \log(N)$ scaling of known sort implementations and selection algorithms look very good in theory, in practice the constant $A$ depends on the speed with which one can sort or rearrange a few numbers.  The computation of the sample quantile $V$ is rather difficult to implement using floating point hardware and it does not vectorize well\footnote{At this point we would like to bemoan the lack of sorting primitives on contemporary CPUs. A similar regret can be voiced for absence of primitives for modular arithmetic.}. What is needed is an algorithm which scales linearly with the number of samples and that can be implemented using floating point instructions.

\section{Derivation of additive upper limit statistic}

In probability theory distribution-independent bounds are commonly obtained by use of Chebyshev-Bienaym\'e's or Markov's inequalities, however they are rarely used in practice, since in common applications they provide bounds that are far too loose.

For example, Encyclopaedia Britanica writes ``Unfortunately, with virtually no restriction on the shape of an underlying distribution, the inequality is so weak as to be virtually useless to anyone looking for a precise statement on the probability of a large deviation. To achieve this goal, people usually try to justify a specific error distribution, such as the normal distribution...'' \cite{brittanica}.

This is because even though Chebyshev-Bienaym\'e's or Markov's inequalities are sharp - turning into equalities for an appropriate probability distribution - these distributions are rarely encountered in practice. 

There exists a stronger Vysochanskij-Petunin inequality \cite{vp_ineql} but it relies on distributions being unimodal - an assumption that is hard to establish in empirical data. A review of other Chebyshev type inequalities can be found in \cite{three_sigma, savage}.

We engineer an upper limit statistic by starting with Markov's inequality
\begin{equation}
P(|X|\ge a)\le \frac{\E(|X|)}{a} 
\end{equation}
and modifying it to read
\begin{equation}
P(|f(X)|\ge a)\le \frac{\E(|f(X)|)}{a}
\end{equation}
Then a further modification yields:
\begin{equation}
P\left(\left|f\left(\frac{\mu-X}{\sigma}\right)\right|\ge a\right)\le \frac{\E\left(\left|f\left(\frac{\mu-X}{\sigma}\right)\right|\right)}{a}
\end{equation}
for, in general, arbitrary $\mu$ and $\sigma>0$ - though in practice these are chosen to be estimates of the mean and standard deviation.
After setting 
\begin{equation}
a=\frac{\E\left(\left|f\left(\frac{\mu-X}{\sigma}\right)\right|\right)}{\epsilon}
\end{equation}
we obtain
\begin{equation}
\label{eqn:unid_ineql}
P\left(\left|f\left(\frac{\mu-X}{\sigma}\right)\right|\ge \frac{\E\left(\left|f\left(\frac{\mu-X}{\sigma}\right)\right|\right)}{\epsilon}\right)\le \epsilon
\end{equation}

Because the original Markov's inequality is correct for a random variable $X$ with an arbitrary distribution, inequality (\ref{eqn:unid_ineql}) is valid for any choice of $f(x)$, $\mu$ and $\sigma$ - even when $\mu$ and $\sigma$ are estimated from the data $X$.

We can now optimize $f(x)$ to provide more precise upper limits or confidence intervals for our desired distribution. 
As a quick example, the inequality \ref{eqn:unid_ineql} becomes sharp for a Gaussian random variable X when we choose $\mu=\E(X)$, $\sigma=\sqrt{\Var X}$ and use a step function
\begin{equation}
f_s(x)=\left\{\begin{array}{ll}
            1 & \textrm{when~} x\ge \hat{x}_\epsilon \\
	    0 & \textrm{otherwise}
            \end{array}
\right.
\end{equation}
where the lower tail cutoff $\hat{x}_\epsilon$ satisfies
\begin{equation}
\label{eqn:xeps}
{\mathcal F}(-\hat{x}_\epsilon)=\frac{1}{\sqrt{2\pi}}\int_{-\infty}^{-\hat{x}_\epsilon} e^{-x^2/2}dx=\epsilon 
\end{equation}

The choice $f(x)=f_s(x)$ is difficult to apply to establish a confidence interval because the function $f_s(x)$ is not invertible: it can happen that the average of $\frac{1}{\epsilon}\left|f\left(\frac{\mu-X}{\sigma}\right)\right|$ for practical data is greater than $1$ which does not yield a constraint on $X$. One approach could be to pick initial $x_\epsilon$ as defined by equation \ref{eqn:xeps} and then iterate to establish a bound for $X$. This is cumbersome for both analytical and numerical computation.

\begin{figure}[htbp]
\includegraphics[width=3.4in]{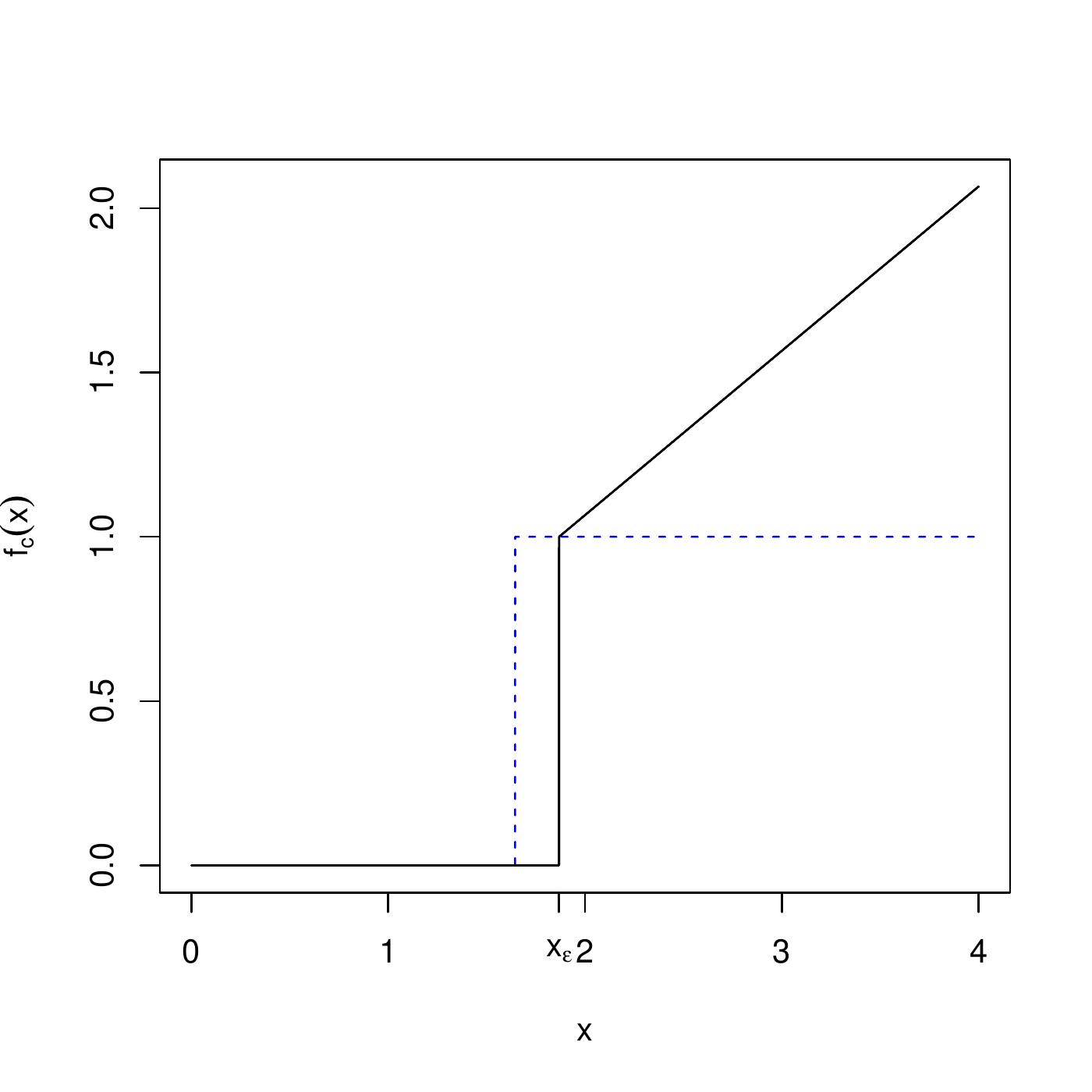}
\caption{Function $f_c$ used for computation of $95\%$ confidence level upper limits (simulation results are shown in figure \ref{fig:distribution_comparison}). The point $x_\epsilon$ has been shifted to the right to compensate for errors in mean estimates for $501$ points of input data. The dashed line shows the step function $f_s$ that makes inequality \ref{eqn:unid_ineql} exact in the ideal case. (color online)}
\label{fig:f_c}
\end{figure}

A better way is to pick the function $f(x)$ to be invertible above $x_\epsilon$. An especially simple and computationally efficient example, shown in figure \ref{fig:f_c}, is given by
\begin{equation}
\label{eqn:f_c}
f_c(x)=\left\{\begin{array}{ll}
            1+\frac{1}{2}\left(x-x_{\epsilon}\right) & \textrm{when~} x\ge x_\epsilon \\
	    0 & \textrm{otherwise}
            \end{array}
\right.
\end{equation}
with the corresponding inverse function given by
\begin{equation}
\label{eqn:f_c_inv}
f_c^{inv}(x)=\left\{\begin{array}{ll}
            x_\epsilon+2(x-1) & \textrm{when~} x\ge 1 \\
	    x_\epsilon & \textrm{otherwise}
            \end{array}
\right.
\end{equation}

Our correction $C$ is then
\begin{equation}
\label{eqn:C}
C=\sigma f_c^{inv}\left(\frac{1}{\epsilon}\E\left(\left|f_c\left(\frac{\mu-X}{\sigma}\right)\right|\right)\right) 
\end{equation}
with expectation replaced by average for empirical data.

\begin{figure}[htbp]
\framebox{
\parbox{3.22in}{
\begin{enumerate}
 \item Prepare by computing value of $$x_\epsilon=-{\mathcal F}^{-1}(\epsilon)+\max(5/\sqrt{N}, \eta)$$
 where $\eta=0.04 \left(\sqrt{\log \frac{N^2}{2\pi}}-{\mathcal F}^{-1}(\epsilon)\right)$
 \item Compute $M=\max_{i=1..N} d_i$
 \item Compute $\mu=\frac{1}{N-1}\left(\sum_{i=1}^N d_i-M\right)$. 
 \item Compute $\sigma=\frac{\sqrt{2 \pi}}{N}\sum_{i=1}^N \max(\mu-d_i, 0)$. We prefer this formula as it is simpler to compute and is insensitive to outliers in the upper tail of the distribution.
 \item Compute $\delta=\frac{1}{N\epsilon}\sum_{i=1}^N f_c\left(\frac{\mu-d_i}{\sigma}\right)$.
 \item Establish upper limit $\textrm{UL}=M-\mu+\sigma f_c^{inv}(\delta)$
\end{enumerate}
}
}
\caption{Algorithm for computing the upper limit from a single batch of data of $N$ points.}
\label{fig:batch_algorithm}
\end{figure}

\section{Implementation of additive upper limit statistic}

A step by step algorithm for computing the upper limit is shown in figure \ref{fig:batch_algorithm}. It incorporates three adjustments that we find important in practical implementation.

First, for smaller values of $N$ the point $x_\epsilon$ has been increased by $5/\sqrt{N}$ compared to theoretical value. This increase guards against fluctuations in the mean estimate $\mu$ which could lead to effectively underestimate $x_\epsilon$. The offset was chosen to correspond to $5$-sigma level in the case of Gaussian variables $\xi_i$, which virtually eliminates underestimate errors.

For large $N$ we provide a fixed offset $\eta$ that provides a consistent overestimate. The somewhat complicated form of $\eta$ is just an approximation of the expectation value of the upper limit times the desired overestimate for large $N$. For extremely large $N$ the term with the square root can be replaced with a more precise (but longer) expression for the expectation of the maximum of $N$ Gaussian variables \cite{extremes}.

Secondly, the mean $\mu$ is computed after excluding the maximum point in the distribution. The exclusion is not necessary if it is known that only weak signals are expected. Otherwise, the overestimate will increase with signal strength. Our implementation in PowerFlux actually excludes a window of 20 frequency bins to each side of maximum, to guard against spillover of detector artifacts due to Doppler shifts.

Lastly, the standard deviation $\sigma$ is estimated using data from the lower tail of the distribution only.
This is reasonable as the upper limit only needs correction for noise values that decrease the maximum of $d_i$ compared to the pure signal case. Also, in practical use, $d_i$ are often bounded from below such as the case of $\chi^2$ distributions resulting from power sums. This leads to slightly smaller upper limits than would be obtained by considering both tails.

The main steps 2-6 of the algorithm \ref{fig:batch_algorithm} employ only piecewise linear functions allowing for very efficient implementation on virtually any computational platform.

\section{Heuristic explanation of the algorithm}

To understand how the algorithm works it is useful to examine steps 5 and 6 in detail. 
First, we convert our data into SNR units:
\begin{equation}
\textrm{SNR}_i=\frac{d_i-\mu}{\sigma}
\end{equation}
Then we compute $\delta$:
\begin{equation}
\delta=\frac{1}{N\epsilon}\sum_{i=1}^N f_c(-\textrm{SNR}_i)
\end{equation}

If we were using the step function then $\delta$ would just be the ratio of the number of data points with SNR larger than $x_\epsilon$ to the number of such points we would expect for Gaussian distribution. Keeping in mind that the step function does not provide a useful constraint for $\delta>1$ we would be, in effect, making a check ``Have we seen too many points in the lower tail ?''. If the answer is yes, we would declare the data non-Gaussian and refuse to set upper limit, if the answer is no, we would use the standard expression $M-\mu+\sigma x_\epsilon$.

By using a piecewise linear function we replace this hard check with a soft correction. We compute $\delta$ as
\begin{equation}
\delta= \frac{1}{N\epsilon}\sum_{\textrm{SNR}_i< -x_\epsilon} 1-\frac{1}{2}SNR_i
\end{equation}
As before, when $\delta\leq 1$ we just use the estimate for the Gaussian: $M-\mu+\sigma x_\epsilon$, but instead of bailing out when $\delta$ exceeds unity we use an inflated estimate:
\begin{equation}
\label{eqn:UL_large}
\textrm{UL}(\delta>1)=M-\mu+\sigma \left(x_\epsilon-2+\frac{1}{N\epsilon}\sum_{\textrm{SNR}_i<-x_\epsilon} 2-SNR_i\right)
\end{equation}
By equation \ref{eqn:unid_ineql} this automatically adjusts our correction $C$ to match the underlying distribution. 

The overestimate can be expressed as:
\begin{equation}
1+\alpha=\frac{P(\delta \leq 1) \left(\bar{M}+x_\epsilon\right)+P(\delta > 1) \expect\left(\textrm{UL}(\delta>1)|\delta >1\right)}{\bar{M}-{\mathcal F}^{-1}(\epsilon)}
\end{equation}
where $\bar{M}=\expect\left(\max \xi_i\right)-\mu\approx \sqrt{\log\left(\frac{N^2}{2\pi}\right)}$.

The overestimate is at least as large as
\begin{equation}
\label{eqn:overestimate}
\alpha\ge \frac{x_\epsilon+ {\mathcal F}^{-1}(\epsilon)}{\bar{M}-{\mathcal F}^{-1}(\epsilon)}
\end{equation}
with the equality approached when $P(\delta > 1)\approx 0$.

To understand the dependence of $P(\delta > 1)$ on $x_\epsilon$ let us consider the case of step function $f_s$, i.e. a simplified $\tilde{\delta}$ variable:
\begin{equation}
\tilde{\delta}= \frac{1}{N\epsilon}\sum_{\textrm{SNR}_i< -x_\epsilon} 1 \le \delta
\end{equation}
The sum is just a binomial variable with parameter $p=1-{\mathcal F}(x_\epsilon)$. Approximating it with a normal distribution for large $N$ we find
\begin{equation}
P(\tilde{\delta}>1)=
1- {\mathcal F}\left(\sqrt{N}\frac{\epsilon-p}{\sqrt{p(1-p)}}\right)
\end{equation}
From this expression it is clear why the gap between $x_\epsilon$ and $-{\mathcal F}^{-1}(\epsilon)$ is essential to good performance: without it the probability of $\delta$ exceeding unity would be more than $50\%$.

The choice of parameters for the algorithm thus involves a tradeoff between narrowing the gap to improve performance for very large $N$ and Gaussian $\xi_i$ and increasing the gap to maintain good performance for small $N$ and arbitrary distributions.

\section{Optimization theory viewpoint}
The algorithm \ref{fig:batch_algorithm} is connected with a convex optimization problem of a linear utility function. Indeed, consider the space of all distribution functions ${\mathcal F}_\rho$ with ${\mathcal F}_\rho(-\infty)=0$ (or, equivalently, measures $d{\mathcal F}_\rho$). The requirement that full probability equals to $1$ is linear:
\begin{equation}
\label{eqn:normalization}
\int_{-\infty}^{\infty} d{\mathcal F}_\rho=1
\end{equation}
and the monotonicity requirement can be expressed as an infinite set of linear inequalities:
\begin{equation}
\label{eqn:monotonicity}
{\mathcal F}_\rho(x_1)\ge {\mathcal F}_\rho(x_2)\textrm{  for all } x_1 \ge x_2
\end{equation}
The steps 5 and 6 of algorithm \ref{fig:batch_algorithm} compute the correction $C$ from the normalized data, thus we can require that our distribution is centered:
\begin{equation}
\label{eqn:centered}
 \int_{-\infty}^{\infty} xd{\mathcal F}_\rho=0
\end{equation}
and normalized, which is achieved either by fixing variance:
\begin{equation}
\label{eqn:variance_condition}
  \int_{-\infty}^{\infty} x^2 d{\mathcal F}_\rho=1
\end{equation}
or by a condition on the lower tail of the distribution as used in step 3 of our algorithm:
\begin{equation}
 \label{eqn:sd_min_condition}
 \int_{-\infty}^{0} xd{\mathcal F}_\rho=-\frac{1}{\sqrt{2\pi}}
\end{equation}
All of these conditions are linear and the distribution functions that satisfy them form a convex set - that is for any two such functions ${\mathcal F}_{\rho_1}$ and ${\mathcal F}_{\rho_2}$ the function $\lambda {\mathcal F}_{\rho_1}+(1-\lambda) {\mathcal F}_{\rho_2}$ also satisfies conditions \ref{eqn:normalization}-\ref{eqn:sd_min_condition} for any $\lambda$ between $0$ and $1$.

If we ignore step 5 and 6 of the algorithm and only use the information above, than the best confidence level we can claim for a given correction value $C_0$ is
\begin{equation}
\label{eqn: utility1}
\epsilon^*(C_0)=\sup_{{\mathcal F}_\rho} \int_{-\infty}^{-C_0} d{\mathcal F}_\rho
\end{equation}
The integral expression that we are to maximize is linear in ${\mathcal F}_\rho$ and thus we have a convex problem with linear utility function. 

We can now apply a well-known fact from optimization theory: if the value of the maximum of linear utility function is reached within a convex set (the domain of our problem) then it will be reached in one of the extremal points of this convex set. This can be easily seen, as any non-extremal point will be inside a linear segment contained in our convex set and one of the ends of the segment should have utility at least as great as the utility of the non-extremal point.

To find the set of extremal points consider that any interior mix  $0<\lambda<1$  of two distributions $\lambda {\mathcal F}_{\rho_1}+(1-\lambda) {\mathcal F}_{\rho_2}$   will have non-zero measure for any set that had a non-zero measure for either ${\mathcal F}_{\rho_1}$ or ${\mathcal F}_{\rho_2}$. A distribution forming extremal point should therefore have the smallest support.

The simplest class of such extremal points satisfying conditions \ref{eqn:normalization}-\ref{eqn:centered}, \ref{eqn:sd_min_condition} is given by Bernoulli measures that yield $-\frac{1}{p\sqrt{2\pi}}$ with probability $p$ and $\frac{1}{(1-p)\sqrt{2\pi}}$ with probability $1-p$.

These are not all extremal points as, for example, these distributions have support that does not approach $0$ closer than $1/\sqrt{2\pi}$. 

However, just by finding the maximum of $\epsilon^*(C_0)$ on these Bernoulli distributions (achieved for $p=1/C_0$) we derive a useful constraint on the estimates one can place with only the knowledge of the mean and standard deviation of our data:
\begin{equation}
\label{eqn:utility1_value}
 \epsilon^*(C_0)\ge \frac{1}{C_0\sqrt{2\pi}}
\end{equation}

\begin{figure}[htbp]
\includegraphics[width=3.4in]{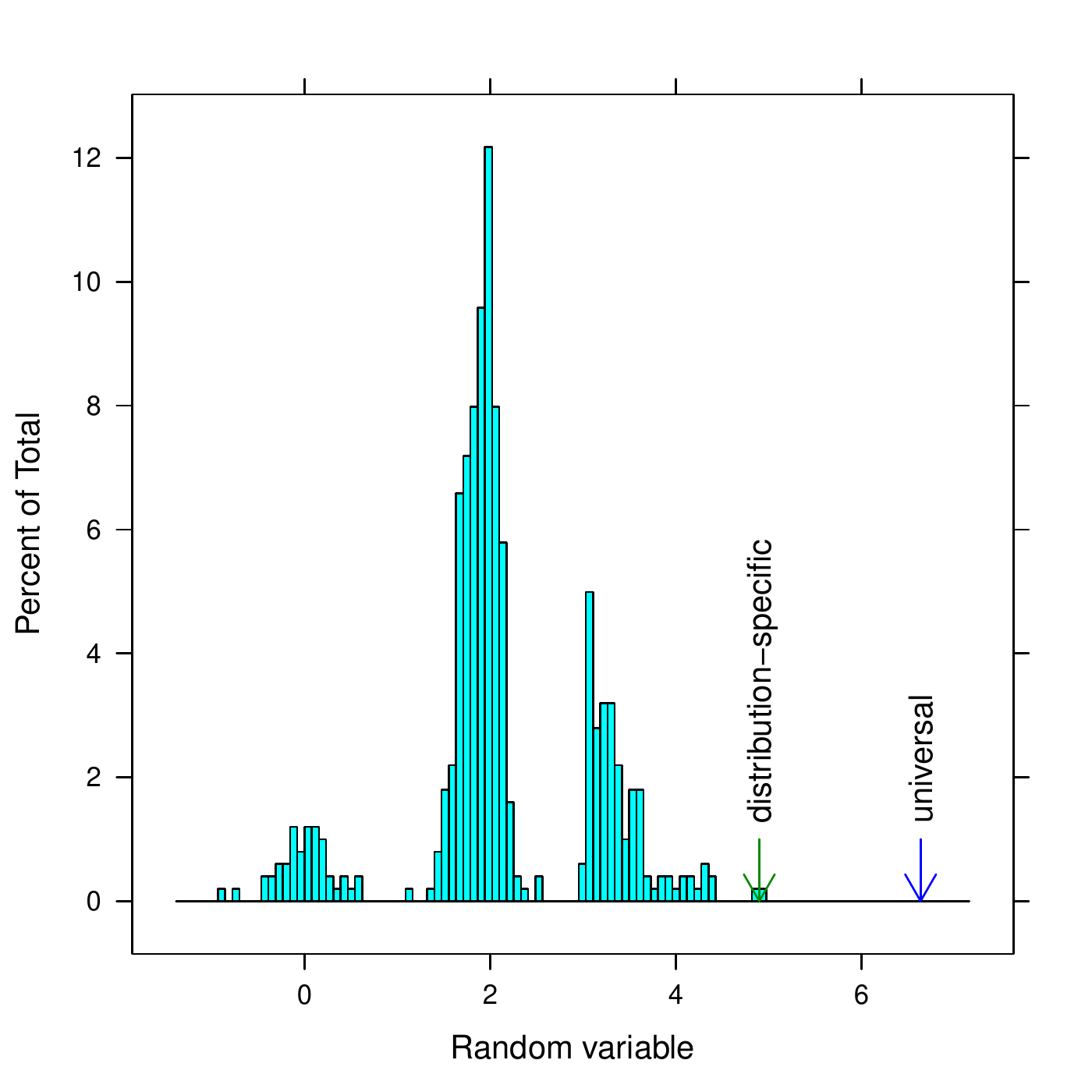}
\caption{Distribution {\tt test1} used in figure \ref{fig:distribution_comparison}. It is composed of three populations, two normal and one exponential. We also show distribution-specific and additive universal 95\% confidence level upper limit for this batch of $N=501$ numbers. (color online)}
\label{fig:test1_distribution}
\end{figure}

This provides a rich source of examples where conventional methods based only on mean and standard deviation would establish an incorrect upper limit. For example, if we pick $\rho$ to be a Bernoulli distribution which yields $0$ in $10\%$ of the cases, our upper limit would have to be at least $10/\sqrt{2\pi}\approx 4$ standard deviations\footnote{This number is for standard deviation computed as in step 3 of our algorithm. For conventional standard deviation a similar computation yields a constraint $C\ge 3$ by considering a $10\%$ distribution and $C>4.35$ by considering Bernoulli distribution with $5\%$ of zeros.} away from $M-\mu$. Yet most conventional methods would use much smaller values, with 3 standard deviations considered very reliable (which it is for Gaussian data). Of course, Bernoulli distribution is a somewhat extreme form of noise, but exactly the same effect is observed in a mix of two Gaussian distributions with different means, effectively smearing out Bernoulli distribution. We illustrate this point with distribution {\tt test1} (figure \ref{fig:test1_distribution}) discussed in the next section.

Let us now turn attention to steps 5 and 6 of the algorithm \ref{fig:batch_algorithm}. The computation of $\delta$ is linear in the distribution functions:
\begin{equation}
\label{eqn:delta}
\delta( {\mathcal F}_\rho)=\int_{-\infty}^{\infty} f_c(-x) d{\mathcal F}_\rho
\end{equation}
We can now let $\epsilon^*$ depend on $\delta$:
\begin{equation}
\label{eqn:utility2}
 \epsilon^*(C_0, \delta_0)=\sup_{\delta({\mathcal F}_\rho)=\delta_0} \int_{-\infty}^{-C_0} d{\mathcal F}_\rho
\end{equation}
and obtain smaller $\epsilon^*$ (and higher confidence level) by slicing our convex domain in level sets of $\delta({\mathcal F}_\rho)$.

Step 6 of the algorithm \ref{fig:batch_algorithm} inverts the relationship by computing correction $C_0$ for a given confidence level and observed $\delta$. As we do not have the exact value of equation \ref{eqn:utility2} the computed correction is somewhat larger, but is still very useful for practical applications as discussed in the next section.

\section{Performance of universal upper limit statistics}

To gauge the performance of universal statistics, we performed a simulation that closely reflects real-world situations we encountered during analysis of LIGO data \cite{FullS5Semicoherent}. 

The PowerFlux search for continuous wave sources iterates over many millions of templates that depend on parameters such as sky location and frequency. As the weights in computed power sums depend strongly on sky location we have to treat each sky location separately. To avoid steep features in frequency spectrum we establish an upper limit for one small frequency range at a time, while holding other parameters fixed. As the possible signal can correspond to any single template the summary data (such as shown on figure \ref{fig:S5_50_200}) is the maximum of individual upper limits over sky and other parameters.

In our simulation we assume that our data consist of independent samples of noise plus a possible deterministic signal in one or more bins. The data are analyzed in batches of $N$  data samples for each of which we establish an upper limit on signal strength. Unless specified otherwise the plots show results for $N=501$, as used in PowerFlux analysis. The final reported value is the worst case (i.e. maximum) upper limit among $L$ batches. We generally expect the performance to improve with the number of batches and the corresponding increase in the signal-to-noise of the loudest outlier. We present plots computed for batch number $L=100$ to model PowerFlux search, as well as batch number $L=1$ which reflects more conventional usage where a single upper limit is established using available data.

\subsection{Upper limit statistics under test}

The plots discussed below show performance of additive universal upper limit statistic, universal statistic based on sample quantiles and three variants of conventional upper limits designed for Gaussian data.

These variants are comprised of ``sd-based'' upper limit which computes sample average $\mu$ and standard deviation. The correction $C$ is then computed as a product of standard deviation and lower $5\%$ quantile of the $t$-distribution.

The second variant - ``modified sd'' upper limit - replaces regular standard deviation with $\sigma$ computed according to step 3 of algorithm \ref{fig:batch_algorithm} and, instead of using a modified $t$-distribution we use $5\%$ quantile of normal distribution, which is expected to work well for large sample sizes.

The third variant - ``mad-based'' upper limit - uses median to estimate $\mu$ and median absolute deviation to estimate $\sigma$. We again use lower $5\%$ quantile of normal distribution.

\begin{figure}[htbp]
\includegraphics[width=3.4in]{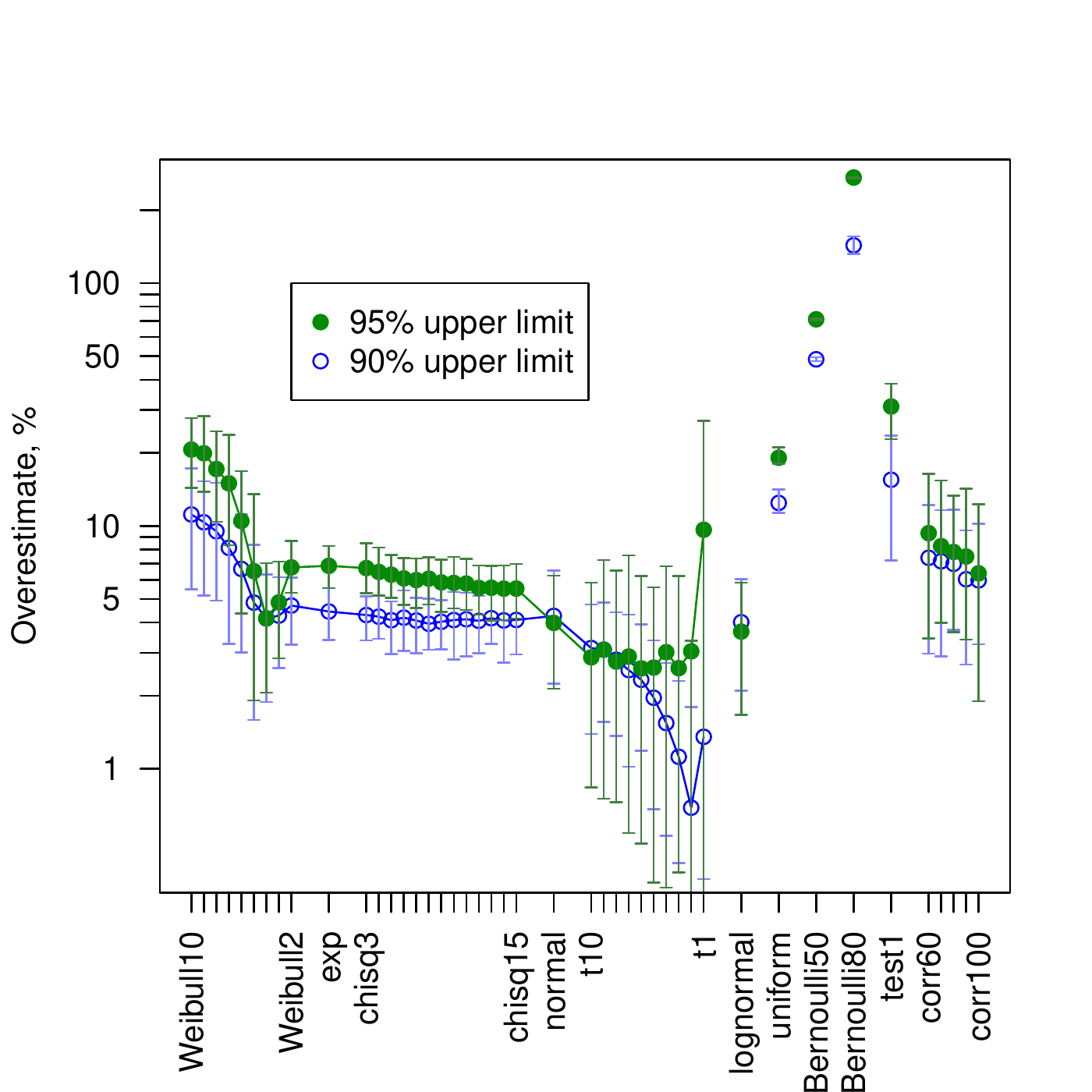}
\caption{Average overestimate of upper limit by the additive universal statistic as compared to the value predicted by analytical formula for the corresponding distributions. The overestimate is less than $5\%$ for Gaussian data, and we expect any practical measurement to perform worse than the ideal case. The error bars show region containing 90\% of upper limits from different noise realizations. The upper limits were computed using $f_c$ (see figure \ref{fig:f_c}) with $N=501$ points of data for different noise distributions described in the text. The points on the graph were obtained by averaging $100$ independent measurements, each of which consisted of finding the maximum among $L=100$ upper limits to simulate maximization across a set of templates. (color online)}
\label{fig:distribution_comparison}
\end{figure}

\begin{figure}[htbp]
\includegraphics[width=3.4in]{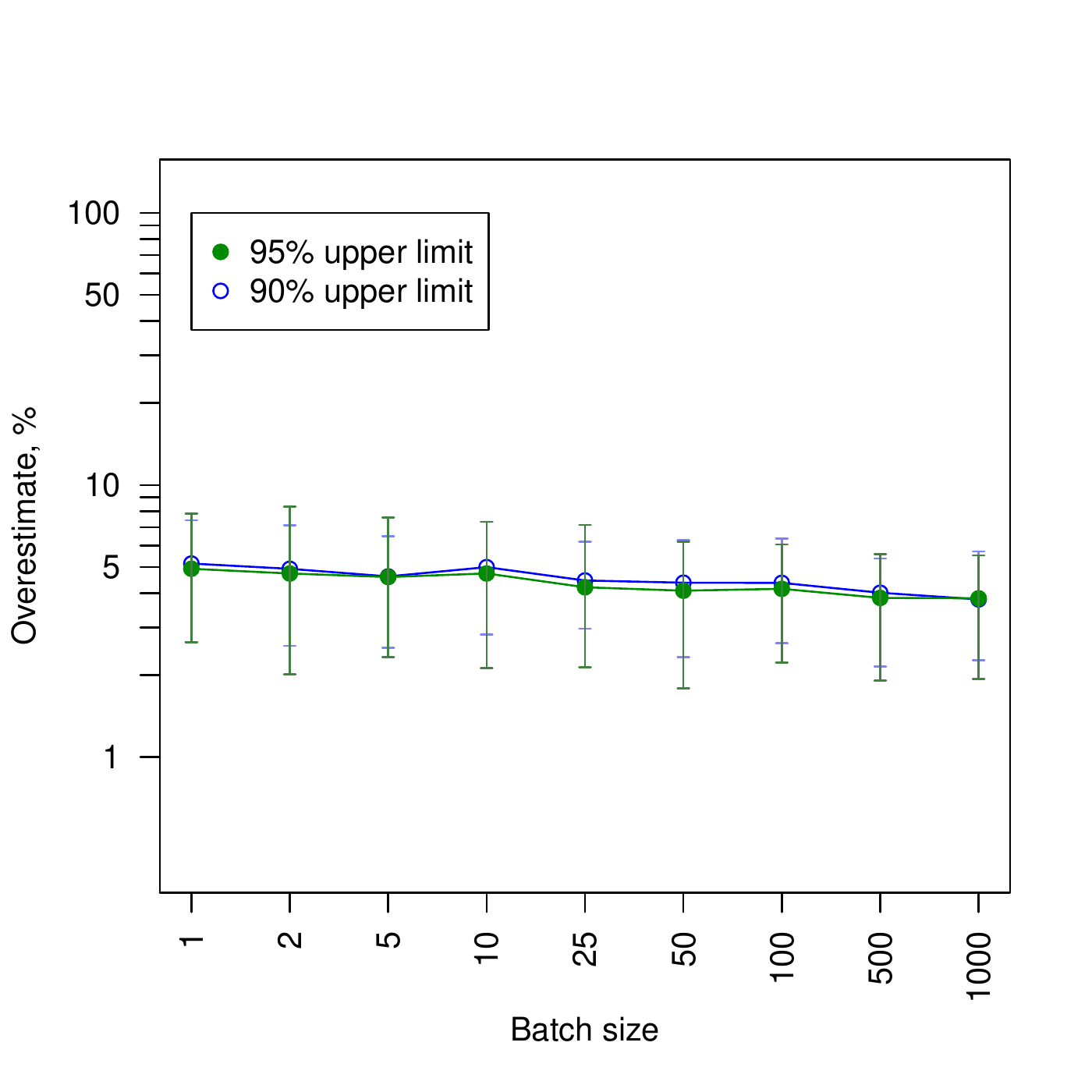}
\caption{Dependence of overestimate of upper limit by the additive universal statistic on the batch size $L$ for Gaussian noise. (color online)}
\label{fig:batch_size_dependence}
\end{figure}

\begin{figure}[htbp]
\includegraphics[width=3.4in]{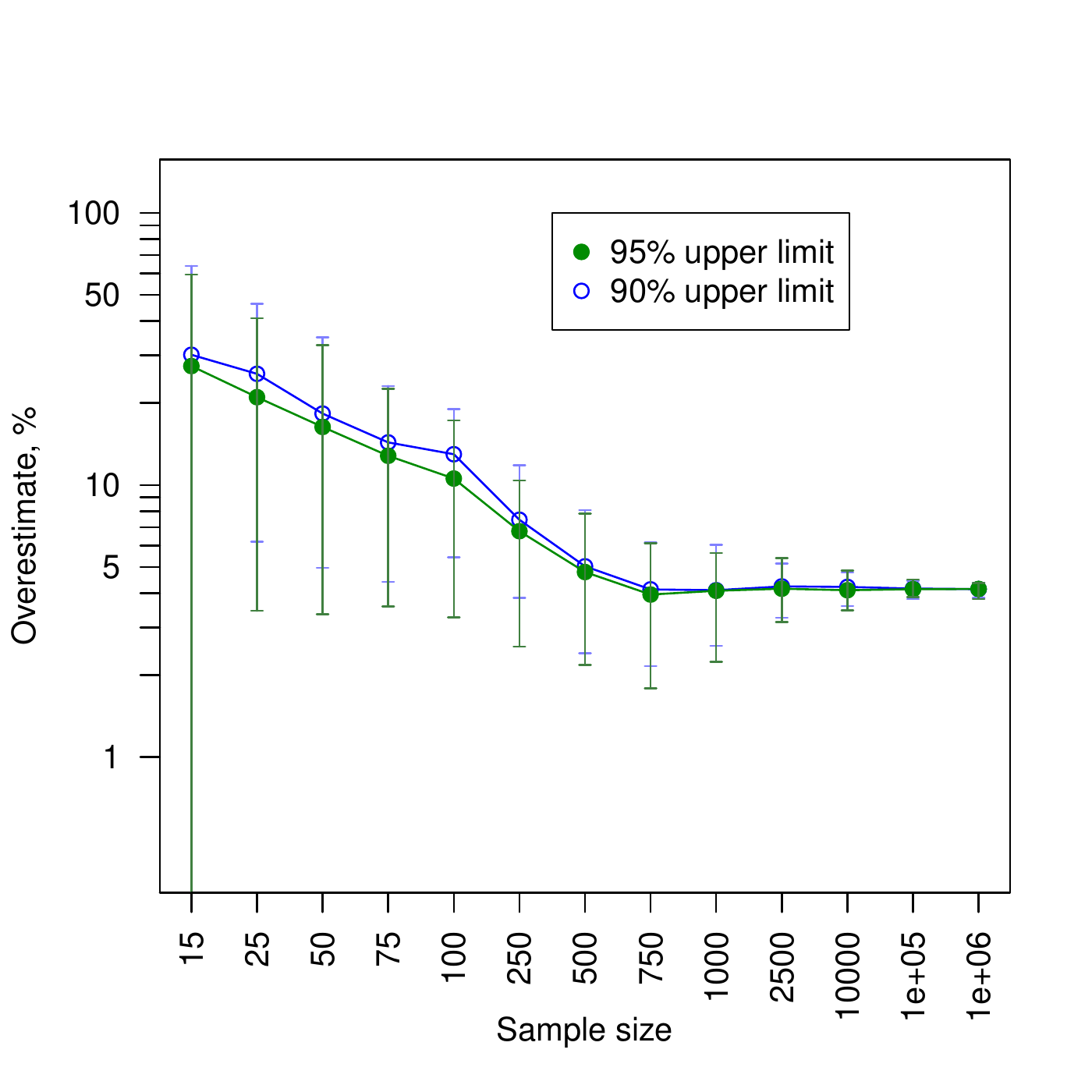}
\caption{Dependence of overestimate of upper limit by the additive universal statistic on the sample size $N$ for Gaussian noise. This plot uses batch size $L=1$. (color online)}
\label{fig:sample_size_dependence}
\end{figure}

\subsection{Noise distributions tested}

To illustrate distribution independent nature of our algorithm we have tested its behavior on a variety of distributions:
\begin{itemize}
 \item Weibull distribution with parameter $k$ with probability density given by:
\begin{equation}
\rho_{\textrm{Weibull}}(x ; k)= k x^{k-1} e^{-x^k}
\end{equation}
The Weibull distributed random variables are always positive. We have included them as a model of power sums with larger lower tail. The plots show results for parameter values from $2$ to $10$.

\item exponential distribution is a special case of Weibull distribution with parameter $k=1$ and $\chi^2$ distribution with parameter $k=2$.

\begin{equation}
\rho_{\textrm{exp}}(x)= e^{-x}
\end{equation}

\item $\chi^2$ distribution has probability density given by:
\begin{equation}
\rho_{\chi^2}(x ;k )=\frac{1}{2^{k/2}\Gamma(k/2)} x^{k/2-1} e^{-x/2}
\end{equation}
It is commonly found as a power distribution (sum of squares) of $k$ independent and identically distributed normal variables. The case $k=2$ coincides with exponential distribution. The limit $k\rightarrow \infty$ gives normal distribution. The plots show data for $\chi^2$ distributions with parameters ranging from $3$ to $15$.

\item normal or Gaussian distributed random variables arise commonly as limits of averages of independent identically distributed variables, via central limit theorem. There are a number of generalizations that relax the assumptions of independence or of having identical distributions, so, in practice, averages of many quantities turn out Gaussian unless there is a specific property that prevents it. Also, Gaussian distribution shows up as the lowest order mode in the quantum harmonic oscillator and optical cavities. 

Normally distributed random variables have probability density:
\begin{equation}
\rho_{\textrm{Gauss}}(x)=\frac{1}{\sqrt{2\pi}} e^{-\frac{1}{2} x^2} 
\end{equation}

\item t-distribution with parameter $k$ has probability density of
\begin{equation}
\rho_{t}(x; k) = \frac{\Gamma\left(\frac{k+1}{2}\right)}{\sqrt{k\pi} \Gamma\left(\frac{k}{2}\right)} \left(1+\frac{x^2}{k}\right)^{-\frac{k+1}{2}}
\end{equation}
This distribution is commonly encountered when the vector of i.i.d. Gaussian variables has been normalized using mean and standard deviation computed from the vector itself. It is an example of a heavy tailed distribution. Interestingly, for parameter values less or equal to $1$ the mean does not exist as corresponding integral is not convergent. The variance only exists for $k>2$. Thus it is particularly interesting to see the performance of various statistics on samples drawn from {\tt t1} and {\tt t2} distributions. The plots show results for $t$ distributions with parameters up to $10$.

\item lognormal distribution is another example of a distribution with heavy tail. Its probability density is given by
\begin{equation}
\rho_{\textrm{lognormal}}(x)=\frac{1}{x\sqrt{2\pi}} e^{-(\ln(x))^2/2}
\end{equation}

\item uniform distribution on segment $[0,1]$

\item Bernoulli distribution with probability $p$ to draw $1$ and $1-p$ to draw $0$. We show performance for $p=0.5$ and $p=0.8$.

\item custom distribution {\tt test1} which is a mix of $10\%$ standard Gaussian variable, $63\%$ Gaussian variable with mean $5$ and standard deviation $0.5$ and $27\%$ exponential variable shifted by $8$ to the right. This reflects a possible case where the power sum is a mix of different distributions. Figure \ref{fig:test1_distribution} shows the histogram of a sample of this distribution.

\item 
A highly correlated Gaussian distribution {\tt corrX} was constructed as 
\begin{equation}
\xi_k^{\textrm{\tt corrX}}=\sum_{j=1}^{X/2} \cos\left(\frac{2\pi k j}{N}\right) \eta_{2j} + \sin\left(\frac{2\pi k j}{N}\right)\eta_{2j+1}
\end{equation}
where $N$ is the size of the noise sample, $k$ is the index of random variable and $\eta_i$ are independent standard Gaussian variables. The results are shown for $X$ from $60$ to $100$ in steps of $10$.

\end{itemize}

\begin{figure}[htbp]
\includegraphics[width=3.4in]{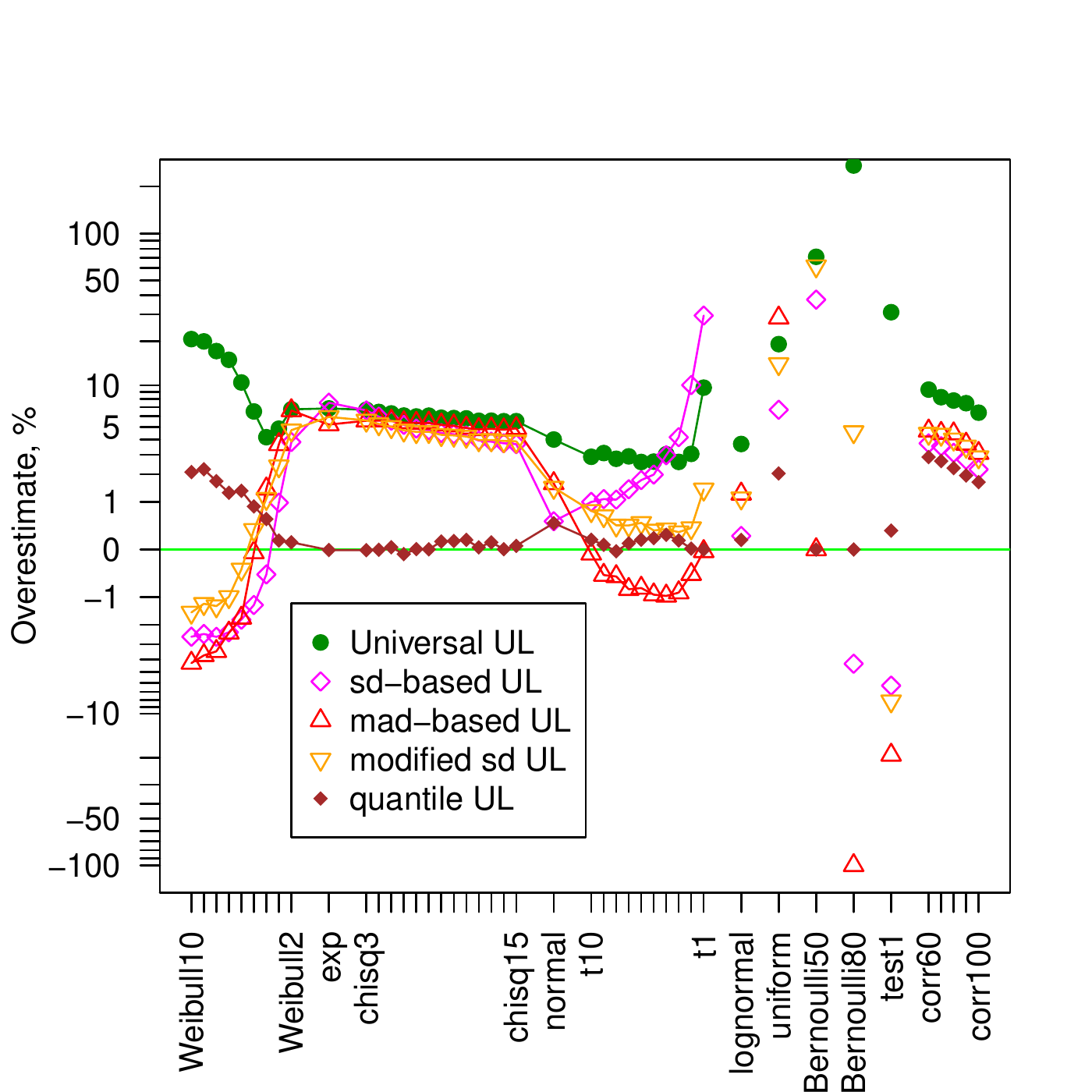}
\caption{Average overestimate of upper limit by various methods as compared to the value predicted by analytical formula for the corresponding distributions. We show 95\% confidence level upper limits computed with the additive universal statistic (solid green dots), a conventional method using Gaussian quantiles and average and standard deviation of the data (magenta diamonds), a variant with modified estimate of standard deviation (orange triangles), a robust variant using median and mad of the data (red triangles) and a quantile based upper limit (solid brown diamonds). The negative values reflect the failure of conventional method based on quantiles for Gaussian distribution to set the correct upper limit in case of some non-Gaussian distributions. This plot shows data for batch size of $L=100$, as described in text. (color online)}
\label{fig:distribution_comparison_to_gaussian}
\end{figure}

\subsection{Upper limit overestimate analysis}

The comparison of this worst case upper limit to the upper limit established from the known distribution of underlying noise is shown in figures \ref{fig:distribution_comparison} and \ref{fig:batch_size_dependence}.  The samples consisted of identically distributed pure noise ($s=0$).

We have averaged the ratios of established upper limits to theoretical ideal values.
As the value of $x_\epsilon$ was obtained assuming Gaussian data we see that our statistic achieves less than $5\%$ overestimate both for $90\%$ and $95\%$ confidence level upper limits. The error bars show region containing $90\%$ of ratios of universal upper limit to theoretical value. The upper mark shows that for Gaussian data we are at or below $7\%$ for $95\%$ of cases. 

A number of other distributions have been tried. As seen on the plot, the performance is remarkably flat for  $\chi^2$ distributions with different degrees of freedom and the overestimate is moderate for uniform distribution. The heavy-tailed Student's t-distributions, as well as lognormal distribution, show good performance as well. In the extreme case of Bernoulli distribution, with $80\%$ probability to obtain $1$, we overestimate by less than a factor of 3 for $95\%$ confidence level. The custom distribution {\tt test1} composed of three populations of normal and exponentially distributed numbers (figure \ref{fig:test1_distribution}) has overestimate of only $31\%$ for $95\%$ confidence level, similar in performance to Weibull distributions or heavily correlated distributions {\tt corrX}.

%

The dependence of overestimate on batch size $L$ is shown on plot \ref{fig:batch_size_dependence} using Gaussian data. The overestimate slowly declines as batch size is increased, but we retain essentially the same performance at all batch sizes.

The dependence on sample size $N$ for the batch size $L=1$ is shown on the plot \ref{fig:sample_size_dependence}. The overestimate is around $30\%$ for $N=15$ and decreases to $4\%$ for very large values of $N$.

\subsection{Comparison of upper limit methods}

In many practical cases the data is assumed to be Gaussian with only a cursory check to its validity. It is thus interesting to compare the behavior of conventional methods on data drawn from non-Gaussian distribution.

We performed the same procedure for upper limits established by conventional methods based on 95\% Gaussian quantiles without any checks that the distribution of data is actually Gaussian. 

\begin{figure}[htbp]
\includegraphics[width=3.4in]{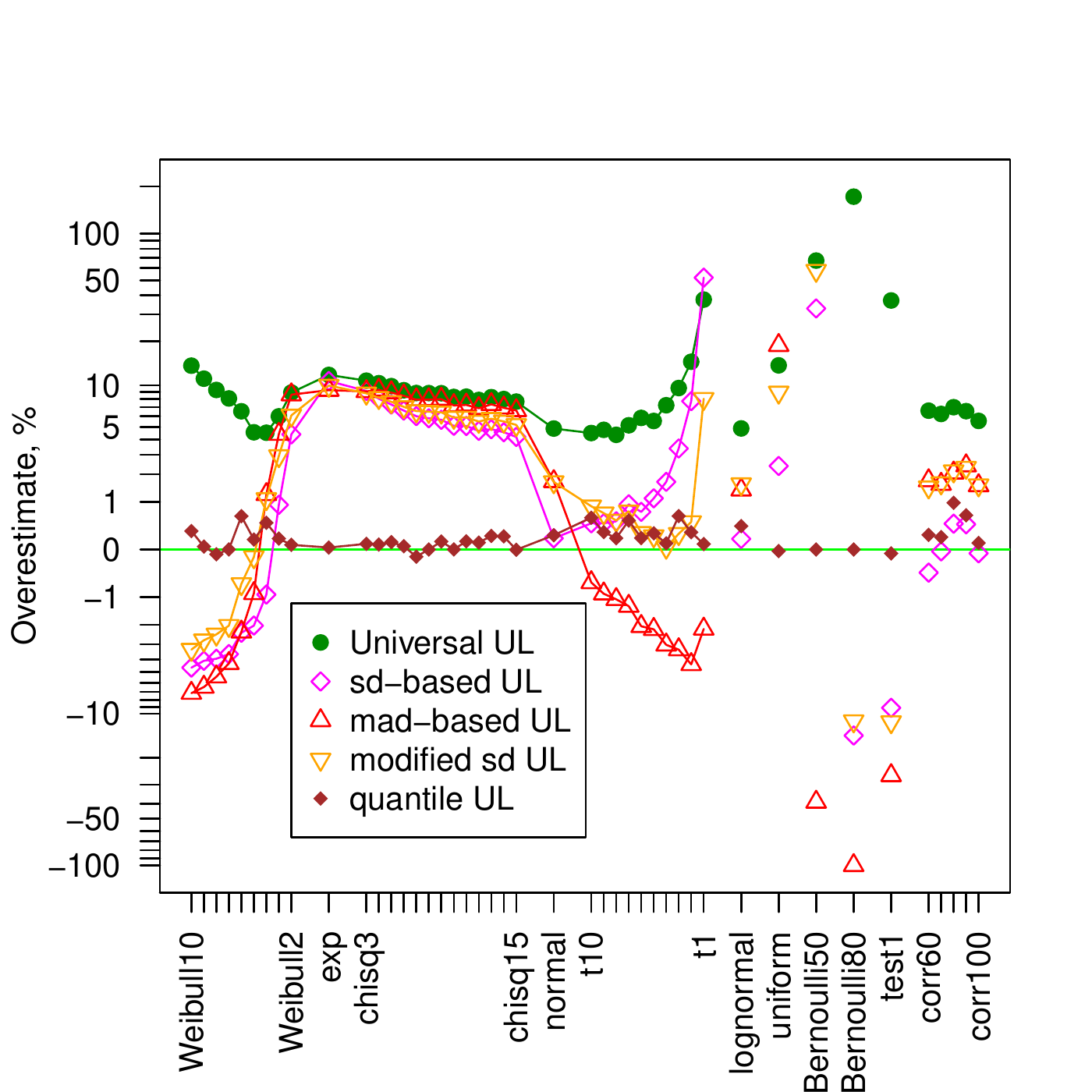}
\caption{Average overestimate of upper limit by various methods as compared to the value predicted by analytical formula for the corresponding distributions. This plots shows 95\% confidence level upper limit for batch size $L=1$ as described in text.
(color online)}
\label{fig:distribution_comparison_to_gaussian_L1}
\end{figure}

The mean and standard deviation of assumed Gaussian distribution were obtained in three ways - through average and standard deviation of the data, average and modified standard deviation as computed by step 4 of algorithm \ref{fig:batch_algorithm} or by using median and median absolute deviation of the data. We also show behavior of the quantile upper limit. The results are shown on figure \ref{fig:distribution_comparison_to_gaussian} for batch size $L=100$ and figure \ref{fig:distribution_comparison_to_gaussian_L1} for batch size $L=1$. 

Interestingly, the mad-based upper limit performs the worst, as it fails to account correctly for the thick tails of Weibull and t-distributions. The mean and standard deviation method fairs better as the presence of a large upper tail increases standard deviation. The modified sd method from figure \ref{fig:batch_algorithm} is more robust and has intermediate performance. For Weibull distributions only the additive or quantile universal statistics return a consistently correct result.

\begin{figure}[htbp]
\includegraphics[width=3.4in]{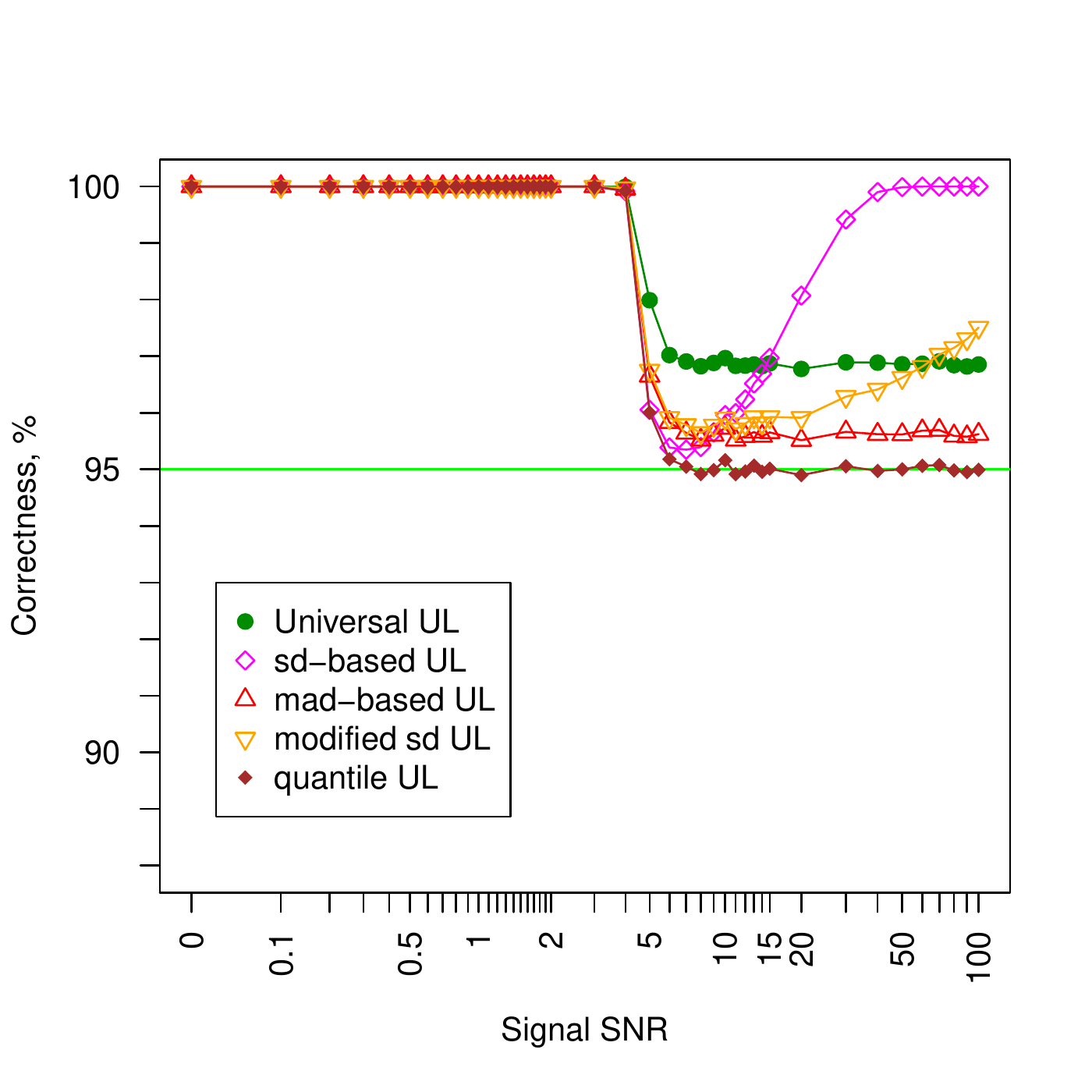}
\caption{Percentage of successfully established upper limits versus injection strength. The noise was distributed according to normal distribution. This plot shows data for batch size of $L=1$. (color online)}
\label{fig:validity_gaussian}
\end{figure}

\begin{figure}[htbp]
\includegraphics[width=3.4in]{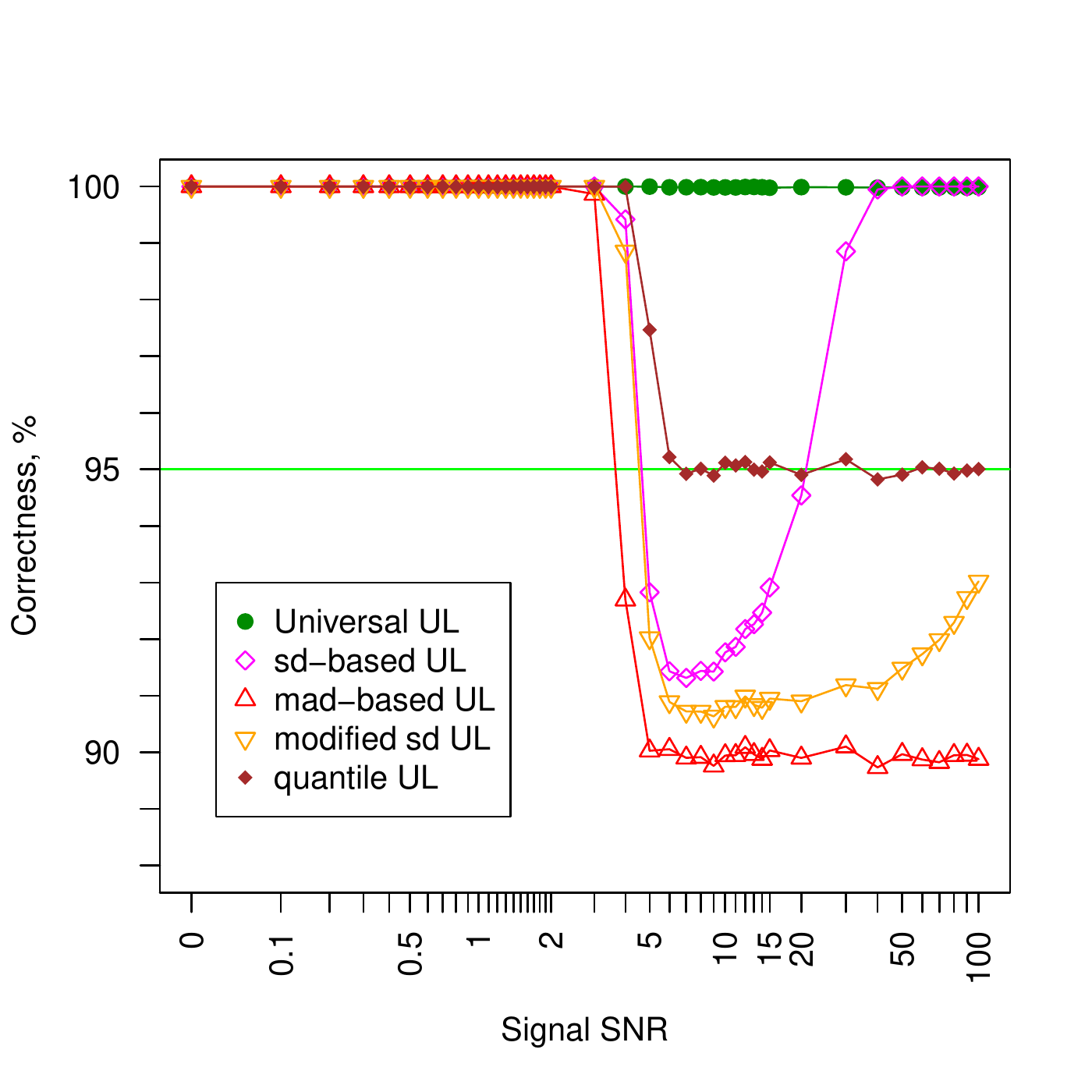}
\caption{Percentage of successfully established upper limits versus injection strength. The noise was distributed according to custom distribution {\tt test1}. This plot shows data for batch size of $L=1$ and sample size $N=501$. (color online)}
\label{fig:validity_test1}
\end{figure}

\begin{figure}[htbp]
\includegraphics[width=3.4in]{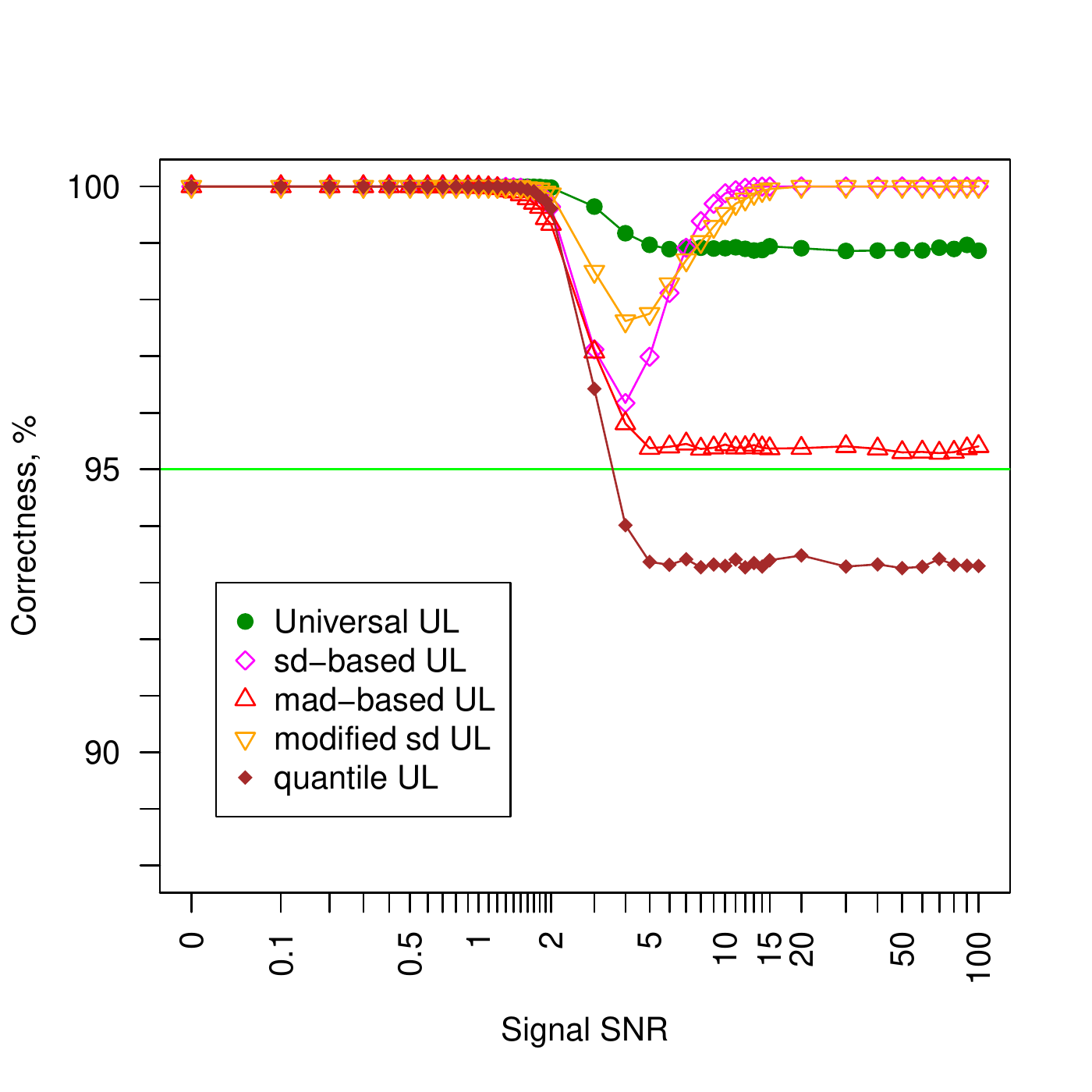}
\caption{Percentage of successfully established upper limits versus injection strength using Gaussian noise.  This plot shows data for batch size of $L=1$ and sample size $N=15$. (color online)}
\label{fig:validity_N15}
\end{figure}

\begin{figure}[htbp]
\includegraphics[width=3.4in]{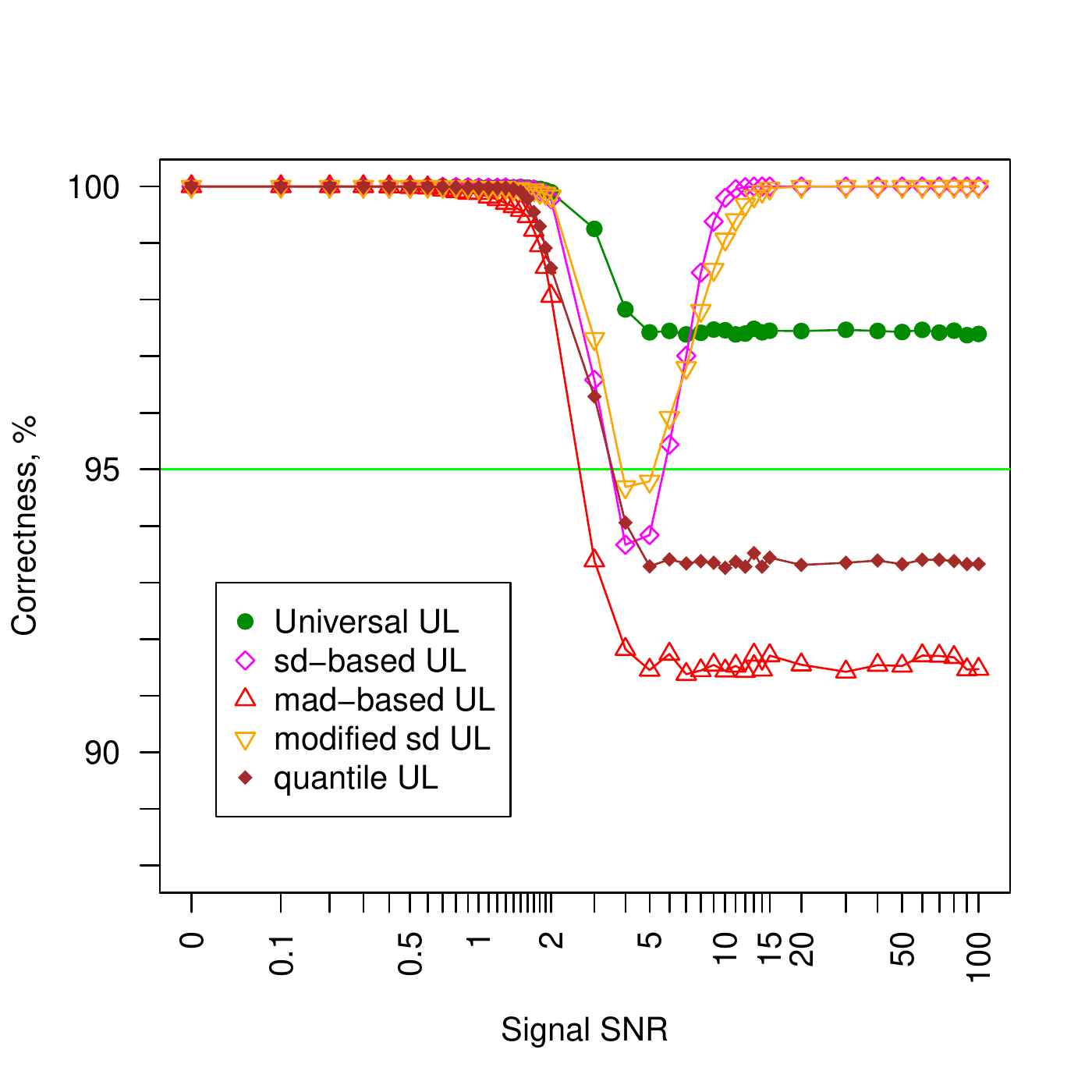}
\caption{Percentage of successfully established upper limits versus injection strength. The noise was distributed according to custom distribution {\tt test1}. This plot shows data for batch size of $L=1$ and sample size $N=15$. (color online)}
\label{fig:validity_test1_N15}
\end{figure}

We study how often the upper limit established by various statistics exceeds injected signal on figures \ref{fig:validity_gaussian}-\ref{fig:validity_test1_N15}. The strength $s$ of injected signal was varied from $0$ (no injection) to $100$ in units of standard deviation of the noise distribution.

 Figure \ref{fig:validity_gaussian} uses Gaussian distributed data while figure \ref{fig:validity_test1} which shows data drawn from {\tt test1} distribution. 


Figures \ref{fig:validity_N15} and \ref{fig:validity_test1_N15} show performance of different upper limit methods for small sample size $N=15$. The quantile based upper limit does not return correct result either for Gaussian noise or {\tt test1} distribution. 

It is interesting to compare performance for Gaussian distribution shown in figures \ref{fig:validity_gaussian} and \ref{fig:validity_N15} to the lower bound given by formula \ref{eqn:overestimate}. For $95\%$ upper limit and $N=501$ we find $x_\epsilon\approx 1.868$ which yields
overestimate for Gaussian data of at least
\begin{equation}
\alpha(N=501)\ge 4.55\% 
\end{equation}
and the expected validity rate of ${\mathcal F}(x_\epsilon)=96.9\%$. This compares well with average overestimate of $\approx 5\%$  (figure \ref{fig:distribution_comparison}) and validity rate of $\approx 97\%$ (figure \ref{fig:validity_gaussian}).

For $N=15$ we compute $x_\epsilon\approx 2.936$ and overestimate bound of
\begin{equation}
\alpha(N=15)\ge 36.5\% 
\end{equation}
and expected validity rate of ${\mathcal F}(x_\epsilon)=99.8\%$. Both numbers are larger than $\alpha\approx 30\%$ (figure \ref{fig:sample_size_dependence}) and actual validity rate of $\approx 99\%$ (figure \ref{fig:validity_N15}). This is unsurprising as our formulas were derived for the limit of large $N$. 

\begin{figure}[htbp]
\includegraphics[width=3.4in]{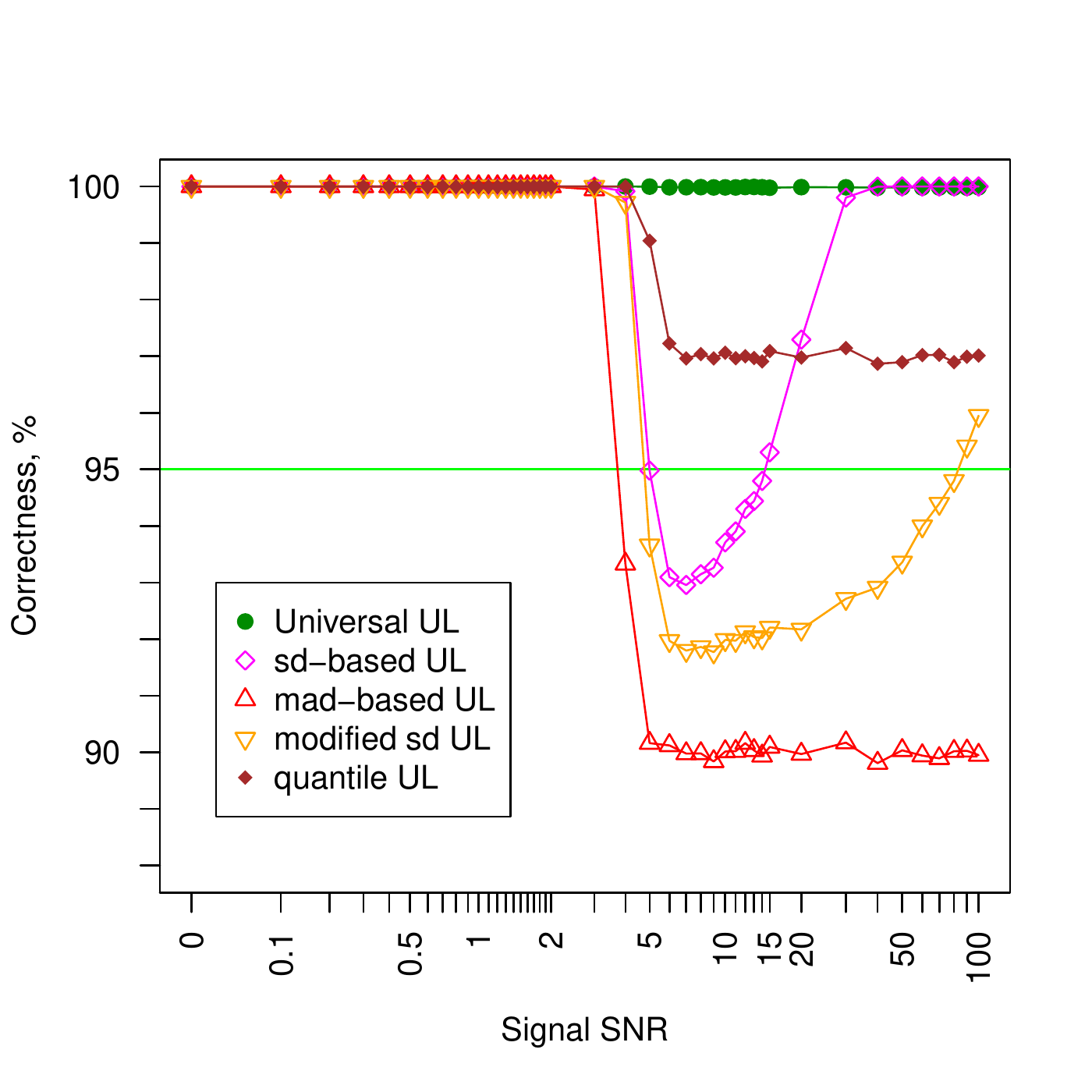}
\caption{Percentage of successfully established upper limits versus injection strength. All statistics, except the additive universal statistic used $96.9\%$ confidence level. The noise was distributed according to custom distribution {\tt test1}. This plot shows data for batch size of $L=1$ and sample size $N=501$. (color online)}
\label{fig:validity_test1_97}
\end{figure}

A natural question to ask is how would the conventional methods perform if we raise the threshold to $96.9\%$ - the confidence level achieved by additive universal statistic on Gaussian data. The results for distribution {\tt test1} are shown on figure \ref{fig:validity_test1_97}. The performance of all statistics has improved, but all statistics based only on estimates of mean and standard deviation still significantly underestimate the confidence level.

\section{Conclusions}
We have described a new {\em universal} statistic that produces reliable and useful upper limits regardless of the underlying distribution of noise, while still producing close to optimum values for a specific family of distributions. We have also shown that conventional methods based only on mean and standard deviation of the sample can significantly underestimate the upper limit. Our Monte-Carlo tests also show that the additive universal statistic is reliable at small sample sizes and with significant correlations in observations.
The algorithm for computing its values is very practical, and is easily implemented for large scale computation. The results of its application to analysis of sixth scientific run of LIGO interferometers are expected to appear in a future paper.

This opens the road for publication of reliable results from large data sets with only partial understanding of distributional properties of data they contain.

\section{Acknowledgments}
This work has been done while being a member of LIGO laboratory, supported by funding from United States National Science Foundation. LIGO was constructed by the California Institute of Technology and
Massachusetts Institute of Technology with funding from the National
Science Foundation and operates under cooperative agreement PHY-0757058.

The author has greatly benefited from suggestions and comments of his colleagues, in particular Evan Goetz, Keith Riles, Alan Weinstein,  and Roy Williams. The exposition was much improved due to suggestions from Reinhard Prix, Sergei Klimenko, Teviet Creighton and two anonymous referees.

This document has LIGO Laboratory document number \texttt{LIGO-P1200065-v8}.


\begin{thebibliography}{99}

\def\etal{{\it et al.}}


\bibitem{FullS5Semicoherent}
B.~Abbott \etal ~ (The LIGO and Virgo Scientific Collaboration),  {\it Phys. Rev. } {\bf D 85}, 022001 (2012)  

\bibitem{brittanica} Chebyshev's inequality. (2012). {\em In Encyclopaedia Britannica}. Retrieved from \url{http://www.britannica.com/EBchecked/topic/108218/Chebyshevs-inequality}.

\bibitem{vp_ineql}
D.~F.~Vysochanskij, Y.~I.~Petunin,  {\it Theory of Probability and Mathematical Statistics} {\bf 21}: 25–36 (1980).

\bibitem{three_sigma}
F.~Pukelsheim, 
{\it The American Statistician}, {\bf 48}, No. 2 (May, 1994), pp. 88-91

\bibitem{savage} 
I.~R.~Savage, {\it Journal of Research of the National Bureau of Standards - B.Mathematics and Mathematical Physics}, {\bf 65 B}, 211-222.

\bibitem{extremes} 
M.~R.~Leadbetter, G.~Lindgren, H.~Rootz\'en, {\em Extremes and related properties of random sequences and processes}, Springer-Verlag (1983)

\end{thebibliography}
\end{document}